\documentclass[3p, twocolumn]{elsarticle}
\usepackage{graphicx}

\def \beq{\begin{equation}}
\def \eeq{\end{equation}}
\def \beqar{\begin{eqnarray}}
\def \eeqar{\end{eqnarray}}
\input{tcilatex}
\begin{document}

\begin{frontmatter}
\title{Phylogenetic Proximity and Nestedness in Mutualistic Ecosystems}
\author[ITBA]{Roberto P.J. Perazzo}
\ead{rperazzo@itba.edu.ar}

\author[CERGY]{Laura Hern\'{a}ndez}
\ead{Laura.Hernandez@u-cergy.fr}

\author[CNEA]{Horacio Ceva}    
\ead{ceva@cnea.gov.ar}

\author[CNEA,CONICET]{Enrique Burgos}
\ead{burgos@cnea.gov.ar}

\author[ITBA]{Jos\'{e} Ignacio Alvarez-Hamelin}
\ead{ihameli@cnet.fi.uba.ar}

\address[ITBA]{Departamento de Investigaci\'{o}n y Desarrollo, Instituto Tecnol\'{o}gico de
Buenos Aires, \\
Avenida E. Madero 399, Buenos Aires, Argentina}
\address[CERGY]{Laboratoire de Physique Th\'eorique et Mod\'elisation, UMR CNRS,
Universit\'e de Cergy-Pontoise,\\ 
2 Avenue Adolphe Chauvin, 95302, Cergy-Pontoise Cedex, France}
\address[CNEA]{Departamento de F{\'{\i}}sica, Comisi{\'o}n Nacional de Energ{\'\i }a At{\'o}%
mica, \\
Avenida del Libertador 8250, 1429 Buenos Aires, Argentina}
\address[CONICET]{Consejo Nacional de Investigaciones Cient\'{\i}ficas y T\'{e}cnicas,\\
Avenida Rivadavia 1917, C1033AAJ, Buenos Aires, Argentina}

\begin{abstract}
We investigate how the pattern of contacts between species in mutualistic
ecosystems is affected by the phylogenetic proximity between the species of
each guild. We develop several theoretical tools to measure that effect and
we use them to examine some real mutualistic sytems. We aim at establishing
the role of such proximity in the emergence of a nested pattern of contacts.
We conclude that although phylogenetic proximity is compatible with
nestedness it can not be claimed to determine \ it. We find that nestedness
can instead be attributed to a general rule by which species tend to behave
as generalists holding contacts with counterparts that already have a large
number of contacts. A nested ecosystem generated by this rule, shows high
phylogenetic diversity. This is to say, the counterparts of species having
similar degrees are not phylogenetic neighbours.
\end{abstract}
\end{frontmatter}

\section{Introduction}

A sustainable management of ecosystems as well as a proper assessment of the
impact of human activity on them can only be achieved with a proper
understanding of the pattern of the interactions between the species. We are
here interested in the case of mutualistic systems. These usually involve
groups of animals and plants, helping each other to fulfill essential
biological functions such as feeding or reproduction. This is the case of
systems in which animals feed from fruits while dispersing the seeds (%
\textit{seed dispersal networks}) or insects feed from the nectar of flowers
while helping the plant in the pollination process (\textit{pollination
networks}).

The structure of such systems is described by means of an \textit{adjacency
matrix} whose elements represent the absence or presence of an interaction
between the plant and animal species. This information when concerning
mutualistic networks strongly indicates that they are not a random
collection of interacting species, but they instead display a high degree of
internal organization. A pervading feature that has been observed is that
the adjacency matrix is nested, i.e. if species are ordered by increasing
number of contacts, those of one species turn out to be nearly a proper
subset of the contacts of the next species in the list \cite{muchas}. This
organization indicates that the \textit{generalists} of both type of species
(i.e. those species that interact with a great number of species of the
other guild) tend to interact among them while there are no contacts among 
\textit{specialists} (i.e. species that interact with very few species of
the other guild).

The nested structure of mutualistic networks has been attributed to a number
of different causes and there is still some degree of controversy about the
ultimate reasons that make this pattern so frequently observed. It is fairly
obvious that a detailed explanation of the interaction behavior of
individual species can be of little help to understand such a generalized
pattern that is found across ecological systems of very different sizes and
types, and involving plants of different nature and animals that range from
insects to birds.

In Ref.\cite{nosotros2} it is claimed that such order may offer some
advantage for the robustness of the whole system thus suggesting that
systems that are currently observed are those that have survived less
disturbed thanks to its nested structure. Other, more elaborated theories
have been proposed. In Ref.\cite{feno} nestedness has been attributed to
phenotypic affinity between species of different guilds while in Ref.\cite%
{Rezende} an extensive analysis is made concluding that phylogenetic
proximity could explain the nested organization of contacts of some cases of
mutualistic systems. A somehow different point of view about this work can
be found in Ref.\cite{comment}.

It has been customary to consider that the occurrence of some positive
statistical correlation is a sign of causation for the occurrence of the
nested pattern of contacts. However, the sole fact that in a part of the
empirical observations two elements appear to be statistically correlated
should not be taken to mean that one is the cause of the other. Such
correlation may rather indicate instead that both elements are not
incompatible, i.e. that they do not mutually exclude each other or that they
stem from a third, common cause. One alternative way to search for causal
relationships is to explore the possible dynamic consequences of some
assumed interaction mechanism, thus validating or falsifying hypotheses
concerning possible interaction mechanisms between the species. In Refs. 
\cite{nosotros1}, \cite{nosotros3} we have proposed a dynamic model, the
Self Organising Network Model (SNM), that allows to study the contact
pattern of a system that is consistent with some hypothetical interaction
mechanism between mutualistic species.

One example of this analysis is given by the strong positive correlation
found between the species' abundance and hence the frequency of
interactions, with the pattern of contacts of some species \cite{Vazquez and
Aizen}. It has been suggested that locally abundant species are prone to
accumulate interactions and conversely rare species are prone to lose them 
\cite{Stang}, as also suggested by neutral theories \cite{natacha}. However,
a dynamic analysis succeeds in generating a realistic distribution of
species degree \textit{under no assumptions} whatsoever on species
abundance. A reverse interpretation may therefore appear also to be
possible, i.e., that accumulation of interactions occurs first, and the
resulting higher reproductive success leads later to local abundance. With
such dynamic modeling and by comparison with the observed natural systems,
it is possible to reject or retain some hypothetical interaction mechanisms
as a possible cause of the observed order.

In considering real ecological systems one should perhaps expect that a
single yet universal cause may be the prime responsible of inducing
nestedness but that other minor causes may help to shape down the observed
pattern of interactions of each individual case \ . This state of affairs
indicates the convenience of finding one or more, model independent tests
that can directly be performed on observational data to gauge the degree in
which an observed pattern of contacts \cite{Lewinsohn} is governed by some
presumed cause.

In the present paper we develop several theoretical tools to measure the
effect of phylogenetic proximity in the contact pattern of the system and
use them to examine some real mutualistic systems. We aim in this way at
establishing whether such proximity can be taken to be responsible for the
emergence of a nested pattern of contacts, while some other minor causes may
explain departures in particular circumstances.

\section{Theoretical Background}

\label{bipartite}

Mutualistic systems can be analyzed as bipartite graphs~\cite{newman}. The
interaction pattern is usually coded into a (rectangular) adjacency matrix
in which rows and columns are labeled respectively by the plant and animal
species. Its elements $K_{p,a} \in\{0,1\}$ represent respectively the
absence or presence of an interaction (contact) between the plant species $p$
and the animal species $a$. The degrees of each plant or animal species can
be obtained from this matrix as:

\begin{eqnarray}
G^{P}(p) &=&\sum_{a} K_{p,a}K^T_{a,p}  \label{grado1} \\
G^{A}(a) &=&\sum_{p}K_{a,p}^{T}K_{p,a}  \label{grado2}
\end{eqnarray}
where $K^T_{a,p}$ is the trasposed of the adjacency matrix.

In order to explore the connection between the phylogenetic structure of
plants and animals and the corresponding pattern of contacts we will assume
that a phylogenetic tree has been established for animals and plants. All
species are therefore taken as the leafs or tips of a tree-like diagram that
represents their evolutionary history.

A prevailing phylogenetic order can be represented as a matrix of
phylogenetic distances between species. In the literature this is obtained
from the expected covariance of traits between pairs of species through a
comparison between real data sets and the results of a random walk along the
tree topology. This way of measuring distances is based on the statistical
model for the transmission of characters along the tree and is therefore
model dependent. Instead of this we use a more direct, unambiguous and model
independent way defining the phylogenetic distance extracting its value from
the topology of the phylogenetic tree as the number of nodes that have to be
explored following the tree and starting from either species, until a common
ancestor of the two species is found. The phylogenetic distance $%
d(k,k^{\prime })$ between any two species $k,k^{\prime }$ is then defined as
the maximum number of nodes that are found starting from $k$ or $k^{\prime }$%
. With this definition the closest distance between any two species is 1. If
all species are at a distance 1 then all species belong to a so-called 
\textit{star-phylogeny}.

This definition is a simplified version of a more elaborate \textit{%
ultrametric} \cite{ultrametricos} distance in which a meaning is ascribed to
the length of the branches of the tree by relating it to the time involved
in the evolutionary history. The distance that we have defined however
satisfies all the requirements of a metric function and it also provides
values that fully agree with an intuitive picture. A small value of $%
d(k,k^{\prime })$ remains associated to species that share the same
branching sequence in a common evolutionary history and a large value
corresponds to species that have been separated at earlier stages.

To test for the effects of phylogenetic proximity in the pattern of contacts
we will consider static and dynamic approaches. Among the former we will
consider the effects of the ecological similarity between species, the
robustness of the model and the direct correlation between the degrees of
pairs of species and their phylogenetic distance. Among the later we will
study the stability of some contact patterns when different rules governing
the interaction between mutualistic counterparts are considered.

\subsection{Static test I: Correlations with the phylogenetic distance}

In order to study the influence of phylogenetic proximity we investigate the
dependence of two parameters with the phylogenetic distance. One is the 
\textit{ecological similarity} $S_{k,k^{\prime }}$ and the other is the
dispersion of the degrees $G(k)$ and $G(k^{\prime })$ of pairs of species $%
k,k^{\prime }$ separated by a given phylogenetic distance.

The ecological similarity is defined~\cite{Rezende} as 
\begin{equation}
S_{k,k^{\prime }}=\frac{W_{k,k^{\prime }}^{A,(P)}}{G(k)+G(k^{\prime })}
\label{ecosimilarity}
\end{equation}%
In (\ref{ecosimilarity}) the matrix $W_{k,k^{\prime }}^{A,(P)}$ corresponds
to the \textit{projected graphs} whose matrix elements measure the number of
mutualistic counterparts that are shared by the species $k$ and $k^{\prime }$%
. The matrix $W$ for plants or animals are obtained through $W^{P}=KK^{T}$
or $W^{A}=K^{T}K$. The parameter \ref{ecosimilarity} measures the number of
mutualistic counterparts shared by the two species as compared to the total
number of counterparts of both, in this way, $S_{k,k^{\prime }|\max }=1/2$.
In order to consider the dependence with the distance we consider the
average $<S(\delta )>$ defined as 
\begin{equation}
<S(\delta )>=\frac{1}{N_{\delta }}\sum_{(k,k^{\prime })|d(k,k^{\prime
})=\delta }S_{k,k^{\prime }}  \label{avecosym}
\end{equation}%
where $N_{\delta }$ is the number of pairs of species of the same guild that
are separated by a phylogenetic distance $\delta $.

If phylogenetically close species share some common feeding or polinization
strategy, they should also be expected to have similar degrees Ref.\cite%
{Rezende}. The second parameter that we consider is therefore the dispersion
of the degrees of pairs of species as a function of the phylogenetic
distance. We define: 
\begin{equation}
\Delta G^{A,(P)}(\delta )=\sqrt{\frac{1}{N_{\delta }}\sum_{(k,k^{\prime
})|d(k,k^{\prime })=\delta }\left( G_{k}^{A,(P)}-G_{k^{\prime
}}^{A,(P)}\right) ^{2}}  \label{deltaG}
\end{equation}%
where $G_{k}$ is the degree of the species $k$. The sum is extended over all
pairs of species $(k,k^{\prime })$ that are separated by the phylogenetic
distance $\delta $.

\subsection{Static test II: Robustness analysis}

\label{robust}

Given the adjacency matrix of a mutualistic system it is possible to get a
measure of the robustness of the system. It has been customary~\cite{Memmot}
to do this through an \textit{attack tolerance curve} (ATC). The basic idea
is that if an ``attack'' is made by which some fraction of the species of
one guild is eliminated, a number of species of the other will become
extinct because they are left without their mutualistic counterparts. It is
of course also possible to determine the species that \textit{survive} to
the above attack.

In order to use this concept to study the effect of the phylogenetic
structure of the species that belong to the mutualist system the ideas
involved in the ATC have to be generalized. Instead of limiting the analysis
to the \textit{number} of species that survive or become extinct, one can
rather study the \textit{phylogenetic relation} between the surviving
species, by keeping track of the average phylogenetic distance between them.

In order to do so we construct a Generalized ATC (GATC) in which nestedness
and phylogeny are simultaneously considered. Nestedness is taken into
consideration by orderly eliminating the species of one guild according to
their degrees while the phylogenetic structure of the partially depleted
counterpart guild is tested through the average phylogenetic distance
between the surviving species.

\subsection{Dynamic test: the SNM}

Several arguments have been put forward to support that the pattern of
interactions between the two guilds of a mutualistic network is due to some
underlying reason. Most of those hypotheses lie upon statistically
significant, observed correlations between some parameters. Correlation is
however not the same as causation. One way out to elucidate a possible
causal link between some hypothesis about the interaction mechanisms of the
species and a given interaction pattern, is to explore eventual consequences
of such hypothesis through a dynamic model.

The basic idea behind this modeling strategy is to verify the \textit{%
consistency} of the empirically observed contact pattern, with some
hypothetical interaction rule that may favor or hamper the contact between
mutualist species. We refer to such interaction mechanism as a contact
preference rule (CPR). An instability may easily be detected when the
observed adjacency matrix tends to be drastically changed if species are
allowed to redefine their contacts to make a better use of some assumed CPR.
Within the same reasoning it is also possible to check the features of the
contact pattern that \textit{emerges} from an initial random adjacency
matrix if iterated changes of this type are allowed to take place. This can
give a clue of what one should expect to observe in nature if some given CPR
is the prevailing interaction mechanism among the species of system. This
kind of analysis allows to trace possible causal relationships between a
generalized individual behavior and the features of the global pattern of
contacts.

From a purely theoretical point of view the above considerations are
equivalent to consider that the observed pattern of interactions corresponds
to an optimal assignment of the contacts between both guilds, with two
constraints, first the fulfillment of some hypothetical, generalized,
individual behavior trait and second, the total energy expenditure of all
the species of the ecological system in their search for contacts. In other
words one may attempt to describe the observed pattern of contacts as the
result of a (combinatorial) \textit{optimization} problem by which contacts
in the adjacency matrix are placed in such a way as to reach an extreme of
some utility function defined in terms of the prevailing CPR.

An example of a dynamic model of this kind is the SNM~\cite{nosotros2}~\cite%
{nosotros1}. This has been introduced in order to account for nestedness in
mutualistic webs. Within that model, the mutualistic system is assimilated
to a bipartite graph and the topology of the corresponding network is
established as the result of a self-organization process. This amounts to
progressively redefine the links obeying some hypothetical CPR that is
assumed to prevail among the nodes of the network. In Refs.\cite{nosotros2},%
\cite{nosotros1} we show that a CPR where species tend to have contacts with
counterparts that already hold a greater number of contacts leads to nested
networks.

The fulfillment of the additional constraint of limiting the total energy
consumed by all the species belonging to the system is built into the model
by imposing that the self organization process takes place keeping constant
the total number of contacts. It is thereforer implicitly assumed that on
the average, the establishment of each contact involves some fixed amount of
energy. One could consider that this is spent in the process of searching
for the counterpart.

It is worth to stress that the self-organization process \textit{does not
represent a real life behavior} of plants and animals of the system. It
therefore does not aim at reproducing any evolutive or adaptive process. It
rather provides a plausible mathematical tool to check for instabilities and
to search for contact patterns that are optimal in the sense explained above.

\section{Results: Tests of the effects of phylogenetic proximity}

We now turn to provide a measure of the influence of the phylogenetic
proximity by focusing our attention in a set of systems that have been
reported in Ref. \cite{Rezende} (NCOR, OLAU, ARR1 and ARR2). Additionally,
we take as a reference an idealized situation through a benchmark system of
comparable size that we call PERF (36 plants and 50 animals). This is
constructed in such a way that its adjacency matrix is perfectly nested and
contacts faithfully follow some assumed (arbitrary) phylogenetic ordering.
The phylogenetic tree and the adjacency matrix are tailored to perfectly
match each other, in the sense proposed in~\cite{Rezende}. The adjaceny
matrix together with the (arbitrary) phylogenetic trees are shown in Fig.\ref%
{Ki}.

In all cases we also consider null models consisting of systems in which the
adjacency matrix and the structure of phylogenetic trees of both guilds are
the same as the observed ones but the species labeling the tips are randomly
permuted. We refer to these as the "randomized" versions of either the real
observed system or the ideal PERF-system. The null model is constructed
considering the average over many realizations of such random sorting. This
is done to provide statistically significant results of a situation in which
the correlation between the structure of the network and the prevailing
phylogenetic order has been destroyed.

\subsection{Correlations with the phylogenetic distance}

\begin{center}
\begin{figure}[tbp]
\includegraphics[width=8cm]{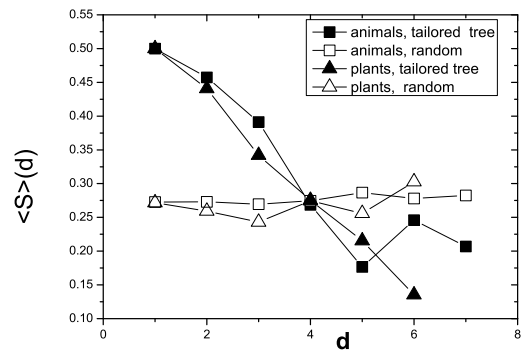} %
\includegraphics[width=8cm]{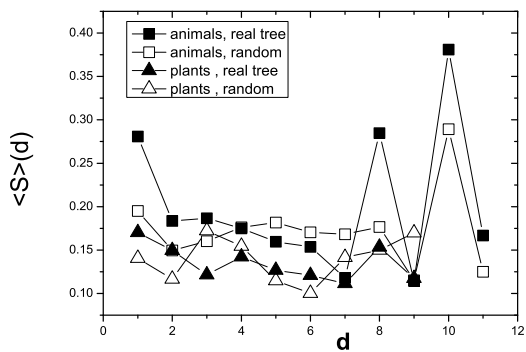}
\caption{Average ecological similarity as a function of the phylogenetic
distance for plants and animals. We show in the same graph the results
corresponding to both, the ordered and the randomized phylogenetic tree.
Upper panel: PERF-system. Bottom panel: NCOR matrix. }
\label{Laura}
\end{figure}
\end{center}

In Fig.\ref{Laura} where we plot the average ecological similarity $%
<S(\delta )>$ defined in Eq.(\ref{avecosym}) as a function of the distance $%
\delta $ separating any pair of species of the NCOR and PERF systems. 
\begin{figure}[tbp]
\vspace{-1.5cm} \includegraphics[width=12cm]{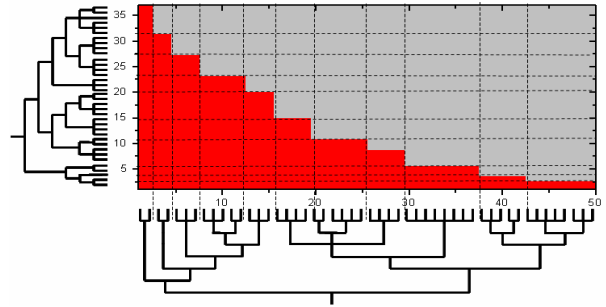} \vspace{-11cm}
\caption{Adjacency matrix of the benchmark PERF system in which a perfectly
nested contact pattern is tailored to follow the (arbitrary) phylogenetic
tree of animals and plants. Dotted lines are drawn to help the eye in the
correspondence of contacts with phylogenetic groups}
\label{Ki}
\end{figure}

It can be seen that for the PERF-system $<S(\delta )>$ decreases
monotonically with the phylogenetic distance for plants and animals. This
monotonous trend is destroyed in the corresponding randomized system. In the
same plots made for the NCOR-system there is no qualitative difference
between the real and the randomized case.

\begin{figure}[tbp]
\includegraphics[width=8cm]{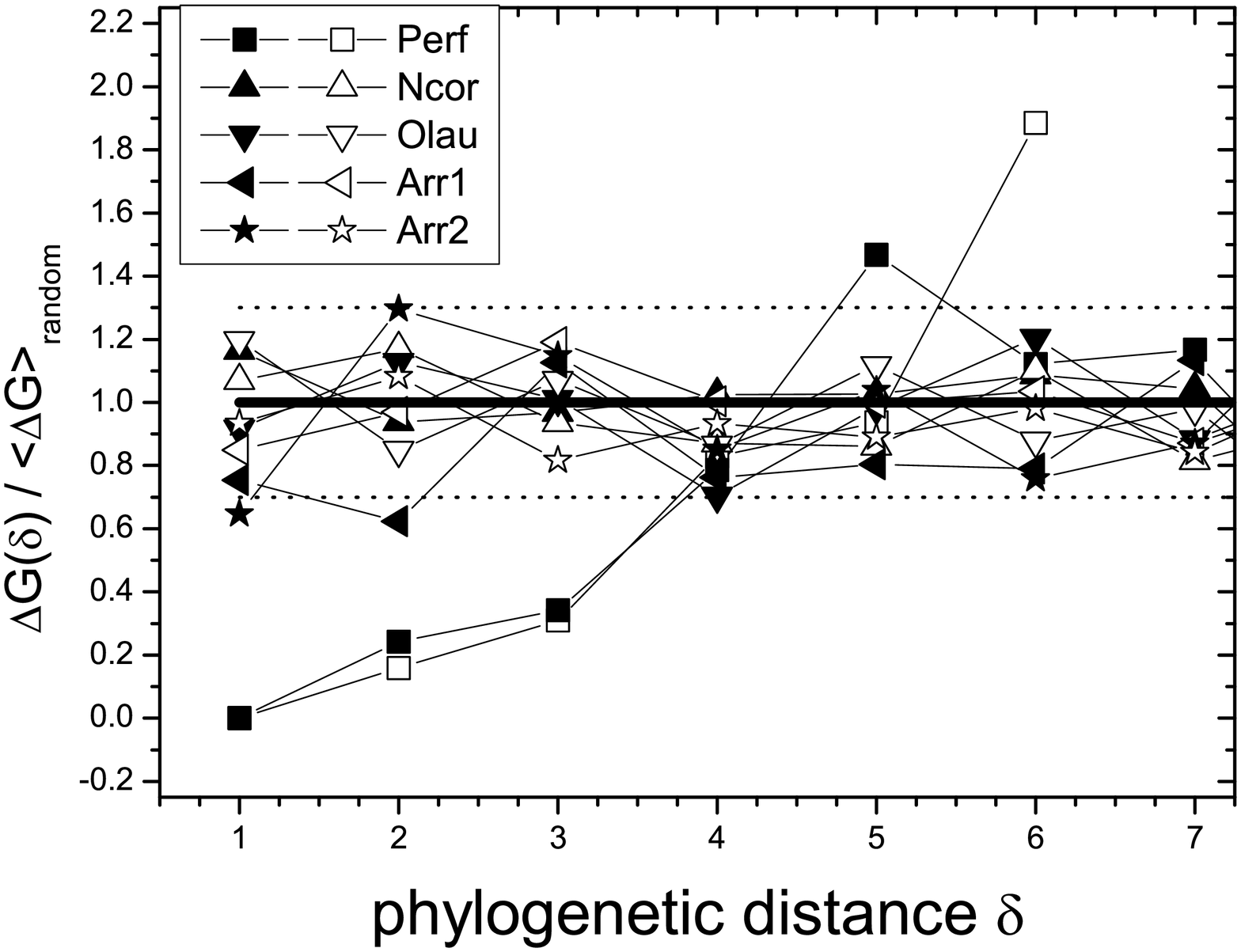}
\caption{$\Delta G$ as a function of the phylogenetic distance $\protect%
\delta$ for $\protect\delta \leq 7$. Filled (empty) symbols represent
animals (plants). In every case, the values are normalized by dividing them
by the average obtained from the null models for the PERF, NCOR, OLAU, ARR1
and ARR2 systems $\langle \Delta G\rangle_{\mbox{random}}$, therefore the
horizontal full line at 1 represents this value.}
\label{Dispersion}
\end{figure}

In Fig.\ref{Dispersion} we show the plots of $\Delta G^{A(P)}(\delta )$ as
defined in Eq.(\ref{deltaG}) for all the systems (NCOR,OLAU, ARR1, ARR2)
together with the one corresponding to the PERF benchmark system. If the
number of contacts were governed by the phylogenetic proximity, the smallest
value of $\Delta G(\delta )$ should be found at $\delta =1$ because
phylogenetically close species must have essentially the same number of
contacts. Since for a greater phylogenetic distance there is also a greater
diversity of species, $\Delta G(\delta )$ must also be an increasing
function of $\delta $. This may no longer hold for very large phylogenetic
distances because of the scarcity of pairs of species in that situation.

Those two features are clearly displayed in the PERF system. Both disappear
when the phylogenetic trees are randomized. The curves derived from the
observed adjacency matrices of the NCOR, OLAU, ARR1 and ARR2 systems \textit{%
without randomization} are compared in Fig.\ref{Dispersion} to an average of
the null models for all systems obtained after 100 randomizations of each
phylogenetic tree. They appear as random positive or negative fluctuations
around the randomized average values thus not showing any appreciable trend
making a difference between the observed and the randomized systems.

\subsection{Generalized Attack Tolerance Curves}

We assume that the system is attacked by eliminating the species of guild A
in decreasing order of their degrees \cite{nota}. In each stage the GATC
that we consider provides the average phylogenetic distance of the fraction
of species of guild B that survive \cite{nota2}. We have found that among a
variety of other possibilities, this is the plot that more dramatically
displays the subtle effects of phylogenetic proximity. The GATC starts at a
value that corresponds to the average distance between the species of guild
B. Since successive stages of the attack produce average distances that are
either the same or smaller, the GATC is supposed to decrease, reaching 0
when only one species of guild B is left.

If contacts were governed by phylogenetic proximity, close neighbors in the
phylogenetic tree of guild A must have similar degrees. Assume that one
starts by eliminating species belonging to one group of phylogenetically
close generalists of guild A. Each elimination stage affects the group of
specialists of guild B that are in turn assumed to be phylogenetic neighbors
(see Fig.\ref{Ki}). Successive extinctions of guild A produces little or no
change in the average distance of survivors of guild B until the moment in
which the last representative of the whole group of specialists of guild B
becomes extinct. The average distance of the surviving species then drops
significantly as a whole \textquotedblleft branch" disappears from the tree.
This fact leads to \textquotedblleft \textit{plateaux}" in the GATC, putting
in evidence the existence of groups of phylogenetically close species of
guild B successively disappearing as the elimination process of guild A
progresses.

This GATC can be compared with a null model in which the tips of the
corresponding phylogenetic trees are randomly sorted and many (100)
realizations of the GATC's are averaged. Upon randomizing any trace of
phylogenetic proximity is eliminated but nestedness is nevertheless
preserved. The GATC is therefore constantly equal to the average distance
for guild B until the moment in which the last species is eliminated and
only one species is left.

\begin{center}
\begin{figure}[tbp]
\includegraphics[width=16cm]{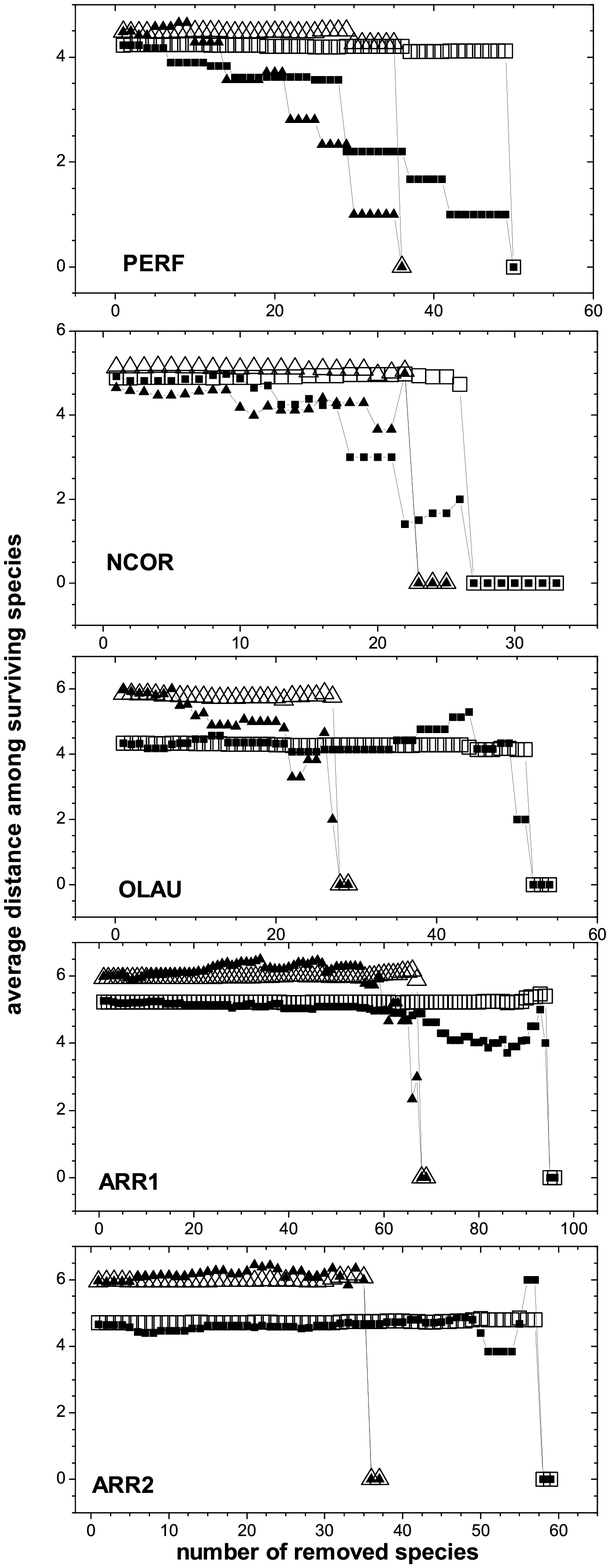} \vspace{-3.0cm} .
\caption{The average phylogenetic distance among surviving plants (animals)
is plot as a function of the number of removed plant (animal) species of the
other guild. Empty symbols correspond to the null model (100 random
realizations of the phylogenetic tree), and filled symbols to the observed
systems. Triangles (squares) correspond to plants (animals)}
\label{atc_1}
\end{figure}
\end{center}

In Fig.\ref{atc_1} we show several plots of the average distance between
surviving species as a function of the number of elimination steps of their
counterparts. In all cases the results of the null model are also drawn for
comparison. The regular plateaux structure can clearly be observed in the
benchmark PERF system. In the GATC for the NCOR system, traces of
phylogenetic affinity can be observed both for plants and animals, leading
to curves that fall well below those of the null model. The GATC for the
OLAU system only displays some structure for plants. In the case of larger
systems ARR1 or ARR2 there is no significant difference between the GATC's
of the real system and those of the null model.

An interesting consequence of this analysis is that the nested structure is
not only robust face to attacks in which concern the preservation of the
ecological system itself but also concerning the preservation of its
ecological diversity. In fact the PERF system looses its diversity as
measured by the decrease of its average phylogenetic distances well before
all real systems of comparable size.

\subsection{A dynamic model}

The influence of phylogenetic proximity can be cast into the form of a
dynamic model of the kind explained previously by properly defining a CPR.
There are two possible algorithms for an SNM fulfilling this. They involve
the following steps.

A plant $p$ is chosen at random and elements of the adjacency matrix $%
K_{p,a_{0}}$ and $K_{p,a_{1}}$ are also chosen at random such that $%
K_{p,a_{0}}=0$ and $K_{p,a_{1}}=1$. Next the following two sums are
calculated: 
\begin{eqnarray}
S_{0} &=&\sum_{p^{\prime }}d(p,p^{\prime })K_{p^{\prime },a_{0}} \\
S_{1} &=&\sum_{p^{\prime }}d(p,p^{\prime })K_{p^{\prime },a_{1}}
\end{eqnarray}%
A swap between this two elements i.e $K_{p,a_{0}}^{new}\rightarrow 1$ and $%
K_{p,a_{1}}^{new}\rightarrow 0$ is proposed and it will be accepted if, and
only if, both animal species $a_{1}$ and $a_{0}$ do not become extinct due
to the swapping.

Moreover, one of the two possible extreme CPR's defining the model is
satisfied:


\begin{itemize}
\item \textbf{MIN:} the condition is $S_{1}\geq S_{0}$

\item \textbf{MAX:} the condition is $S_{1}\leq S_{0}$ 
\end{itemize}

If either of the two conditions is not met, the proposed swapping is
rejected. In order to proceed iteratively, all the above steps have to be
repeated by interchanging in each step the role of plants and animals.

Within the MIN-CPR possibility, the animal counterpart that will finally be
selected for $p$ is such that all other plants $p^{\prime}$ that have some
contact with it are \textit{phylogenetically closer} to $p$. This is so
because the sums $S_0$ and $S_1$ involve all phylogenetic distances between
the plant $p$ that has been selected at random and all other plants that
make some contact with the two animals $a_0$ and $a_1$. With this algorithm
the configuration of contacts is progressively dominated by phylogenetic
proximity: species of one guild are assumed to interact in the same fashion
as the rest of species of the same guild belonging to their phylogenetic
neighborhood.

Within the MAX-CPR possibility, the animal counterpart that will be selected
for the plant $p$ is such that all other plants $p^{\prime }$ that have some
contact with it are more \textit{phylogenetically distant} to $p$. In this
way the set of species that share contacts with $p$ tends to have a greater
phylogenetic diversity. In this alternative all species tend to be as
generalist as possible as far their phylogenetic classifications are
concerned. This alternative bears a greater similarity with the CPR
considered in \cite{nosotros1}. This is because a larger sum $S_{1}$ or $%
S_{0}$ is obtained not only by involving species that are more
phylogenetically distant but also by involving a larger number of
counterparts.

In order to check the ordering process generated by the SNM we define an
effective distance between interacting species through: 
\begin{equation}
D^{A,(P)}=\frac{1}{\langle d \rangle}\frac{\sum_{k,k^{\prime }}d(k,k^{\prime
})\tilde{W}_{k,k^{\prime }}}{\sum_{k,k^{\prime }}\tilde{W}_{k,k^{\prime }}}
\label{paramD}
\end{equation}
In Eq.(\ref{paramD}) $\tilde{W}_{k,k^{\prime }}$ represents the unweighted
adjacency matrix of the projected graphs for animals or plants. Its matrix
elements are 1 (0) if two species share (do not share) mutualistic
counterparts. This equation provides different results for plants or animals
and should therefore be evaluated separately for the two guilds.

The sum in the denominator of Eq.(\ref{paramD}) is just the number of terms
appearing in the numerator, therefore $D$ represents the average
phylogenetic distance between species of the same guild that share at least
one counterpart of the other guild. $D$ is normalized by dividing it by the
average phylogenetic distance $\langle d \rangle$ between \textit{all} plant
(animal) species of the system, namely 
\begin{equation}
\langle d \rangle=\frac{\sum_{k,k^{\prime }}d(k,k^{\prime })}{%
N_{P,A}(N_{P,A}-1)}
\end{equation}
where $N_{P,A}$ is the number of plant or animal species of the system.

A value $D^{P,(A)}<1$ indicates that phylogenetic proximity is a dominant
effect, while $D^{P,(A)}\simeq 1$ is a signature that it is not relevant. We
use its values to check for the convergence of the ordering process implied
in the SNM.

\begin{figure}[tbp]
\includegraphics[width=9.5cm]{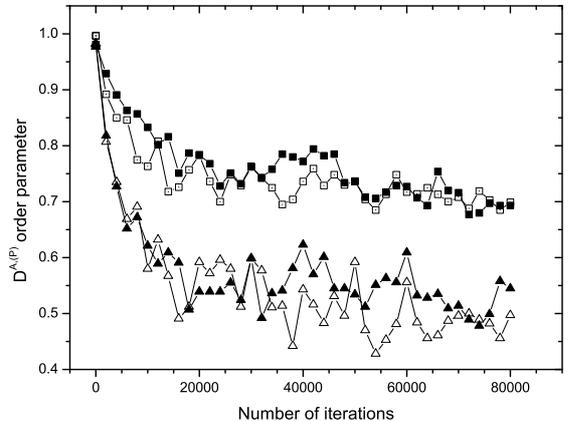}
\caption{Order parameter of Eq.(\protect\ref{paramD}) $D^{P,A}$ as a
function of the number of iteration steps of the SNM algorithm, using the
MIN-CPR alternative. Filled simbols correspond to using as input to the SNM,
the observed data of the NCOR system while empty symbols correspond to using
as input, a random adjacency matrix with the same density of contacts as the
one observed in nature. Triangles (squares) correspond to plants (animals).}
\label{parorden1}
\end{figure}

\begin{figure}[tbp]
\includegraphics[width=9.5cm]{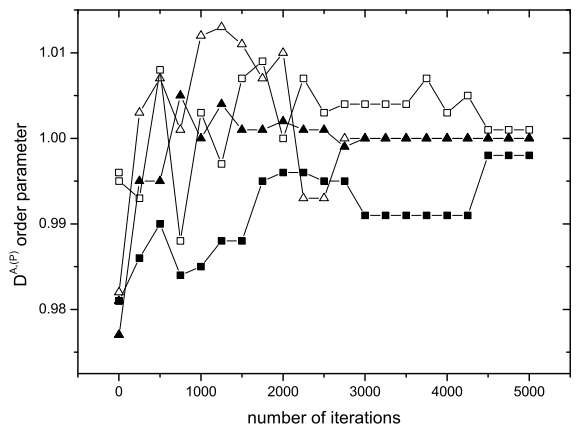}
\caption{Value of the order parameter of Eq.(\protect\ref{paramD}) $D^{P,A}$
as a function of the number of iteration steps of the SNM algorithm using
the MAX-CPR alternative. Notice that the vertical scale is strongly expanded
as compared to that of Fig.\protect\ref{parorden1}. Conventions are the same
as in that figure. }
\label{parorden2}
\end{figure}

In Figs.\ref{parorden1} and \ref{parorden2} we plot the values of $D^{P,(A)}$
as a function of the number of iteration steps of the SNM for plants and
animals for the two possible CPR's considered above. Concerning observed
systems we only consider the NCOR system that is representative of a
possible correlation between nestedness and phylogenetic proximity.

Two different initial conditions are plotted in each graph: one corresponds
to the adjacency matrix has been reported from observation campaigns\cite%
{Rezende}. The other consists in an adjacency matrix of the same number of
rows and columns and with the same number of contacts except for the fact
that they are randomly distributed.

The initial values of $D^{P(A)}$ are very similar for both initial
conditions. These are slightly below the average phylogenetic distances
between species belonging to either of the two guilds. As the number of SNM
iterations grows both CPR's produce values of $D^{P(A)}$ that reach
asymptotic constant values that are essentially independent of the initial
conditions, thus indicating that the system has achieved a stable and
ordered pattern of contacts. However, while for the MIN-CPR case it is found
that $D^{P(A)}$ stabilize at values that are significantly smaller than
unity, for the MAX-CPR alternative it is found that they reach a stable
value of 1.

The initial values of $D^{P,(A)}$ indicate that species of the same guild
sharing at least one mutualistic counterpart, are separated by an average
distance that is very similar to the average distance of both phylogenetic
trees. The fact that the observed value of $D^{P(A)}$ is close to the one
obtained with a random matrix suggests that phylogenetic proximity may be
regarded as an accidental situation involving few species of the system.
While the MIN-CPR favors a contact pattern in which \textit{both} guilds
preserve their phylogenetic proximity, the MAX-CPR favors instead a more
diversified situation in which species hold contacts with a phylogenetically
diverse community.

\begin{center}
\begin{figure*}[tbp]
\includegraphics [width=18cm]{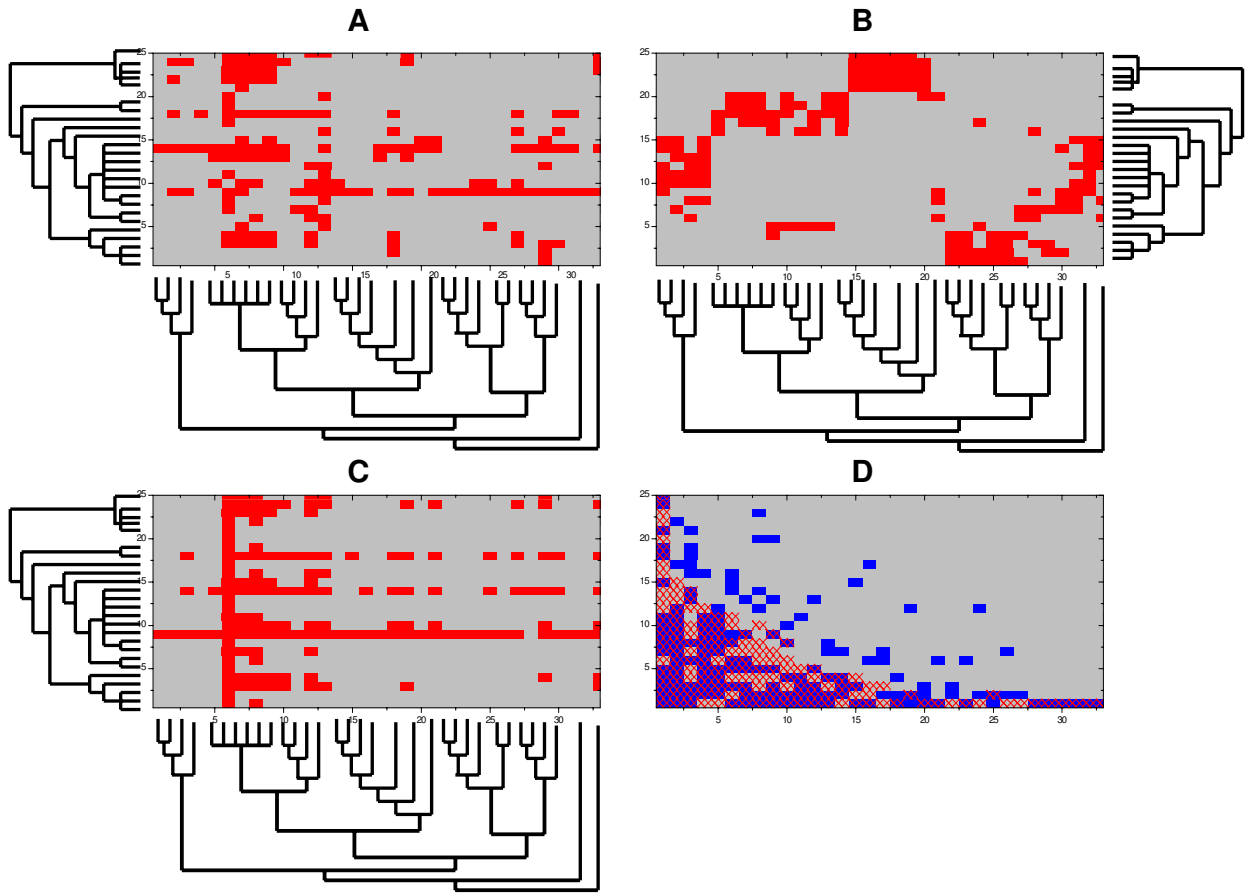}
\caption{Several adjacency matrices of the NCOR system. Panel(A) the
empirical contact pattern with species ordered according to the phylogenetic
tree (shown along both margins of the matrix); panel(B): contact pattern
produced by the SNM after 80,000 iterations using as input the empirical
matrix shown in panel (A) and the MIN-CPR alternative, species in the same
order as in panel (A); panel (C): contact pattern porduced by the SNM after
5000 iterations, using MAX-CPR, species in the same order as in panel (A);
panel (D) same contact pattern as panel (A) (dark pixels) and (C) (hatched
pixels) but species are ordered by their degree. Dark pixels correspond to
observed contacts (Panel (A)) while slanted pattern corresponds to the
theoretical results (panel (C))}
\label{adyacencia}
\end{figure*}
\end{center}

The SNM algorithm can also be used to test if some given pattern of contacts
is compatible with a CPR involving some kind of phylogenetic dependence. The
asymptotic contact pattern can give a clue of what kind of pattern one
should expect to find for each prevailing CPR. In Fig.\ref{adyacencia} we
show the adjacency matrix of the SNM system as obtained after a great number
of iterations of the SNM (panels (B) and (C)). These are the asymptotic
contact patterns corresponding to the MIN-CPR alternative (panel B) and to
the MAX-CPR alternative (panel C). In all cases the initial conditions are
the empirically observed contact pattern (shown in Panel (A)). These
matrices correspond to configurations in which $D^{P,(A)}$ have reached an
almost stationary value and are therefore nearly optimal in the sense
explained above. In the same figure the phylogenetic tree of plants and
animals are shown to guide the eye.

The MIN-CPR corresponds to a rule in which the search of contacts is
dominated by phylogenetic proximity. To better understand the emerging
contact pattern (Panel (B)), one has to bear in mind that both animals and
plants are considered on equal footing. This gives rise to an adjacency
matrix that breaks in disconnected blocks in which phylogenetically close
groups of one guild interact with similar groups of the other guild. This is
the opposite of a nested scheme since upon a greater number of iterations of
the SNM each species tends to specialize its contacts as much as possible.
By the same token, generalists are ruled out of the system. The contact
pattern of the NCOR system used as an initial condition, becomes therefore
severely disturbed putting in evidence that it is unstable under the
presence of the MIN-CPR in which phylogenetic proximity is the dominant rule.

A similar analysis for the MAX-CPR situation shows an opposite behavior. The
SNM causes no drastic reorderings, reinforcing instead the presence of
generalists and keeping the matrix mildly changed (see panel C). The
observed adjacency matrix must therefore be considered stable under such
CPR. This run of the SNM also provides additional information. The NCOR
system hosts a group of animals that are phylogenetically close and that are
all fairly good generalists (the \textit{turdus} group). Such partial
correlation between degree and phylogenetic proximity is not destroyed by
the perturbations introduced by the SNM, if the prevailing CPR is of the MAX
type (compare panels (A) and (C)).

An additional effect of the iteration of the MAX-CPR rule is that it leads
to an asymptotically stable contact pattern that is almost perfectly nested.
In panel (D) we show the adjacency matrices of panels (A) and (C) in which
species have been reordered according to increasing degree, thus comparing
the empirically observed nested structure of the NCOR system with an
asymptotically nearly perfect nested pattern produced by the SNM using the
MAX-CPR. Since the observed NCOR system is considerably nested, the effects
of the SNM are not drastic.

The occurrence of phylogenetically close species with similar degrees and
the occurrence of a nested pattern should therefore be considered as
independent from each other. When speaking of a \textit{cause} for
nestedness one should expect an element that is present in all the observed
systems, with perhaps minor variations in few individual cases. We have
found here that the only general cause of nestedness is essentially the same
as the one reported in \cite{nosotros1}, i.e. a CPR that tends to place
contacts on mutualist counterparts that already concentrate a large number
of contacts. In the present framework nestedness emerges as a consequence of
holding contacts with \textit{phylogenetically diverse groups} of species.
From this point of view, phylogenetic proximity has therefore to be
considered as \textit{compatible} with a rule inducing nestedenss but is far
from being a cause of it.

\section{Conclusions}

We have tried several tests to gauge the influence of phylogenetic proximity
in the occurrence of nestedness. They were performed on several real systems
whose adjacency matrices are reported in the literature (NCOR, OLAU, ARR1
and ARR2 in Ref.\cite{Rezende}) and also on PERF, a benchmark of an ideal
system in which degrees and phylogenetic proximity have been tailored to
display a very strong correlation.

In these tests we have considered the ecological similarity between species,
the dispersion in the degrees of species as a function of the phylogenetic
distances separating them and the shape of the generalized attack tolerance
curves (GATC). In all cases we have also compared the results obtained from
real or ideal systems with those of null models constructed by randomly
sorting the tips of the phylogenetic trees.

The dependence of the ecological similarity or the dispersion of the degrees
with the phylogenetic distance have proven to be essentially
undistinguishable from the one obtained for the randomized systems in all
cases considered. The ideal, perfectly organized system shows regular trends
of both parameters as a function of the phylogenetic distance that are not
met by real systems. One must however bear in mind that both tests are
somewhat \textquotedblleft coarse grained", in the sense that they involve
the adjacency matrix as a whole.

To circumvent this problem we define a GATC in which we plot the average
distance between the fraction of species of one guild that survive after
increasing fractions of the counterpart guild are eliminated. The effect of
phylogenetic proximity reveals itself as the occurrence of plateaux in an
otherwise monotonously decreasing function. The results indicate that the
influence of phylogenetic proximity is more an accident than a rule. It can
only be detected in the NCOR system while in the rest (OLAU, ARR1 and ARR2)
only traces can sometime be observed. Moreover the results indicate that the
nested structure of the network makes the system robust in terms of
ecological diversity. In fact, considering the average phylogenetic distance
as a measure of the ecological diversity, real systems preserve this better
than the PERF system.

In order to investigate for possible causes of nestedness rooted in
phylogenetic proximity we have complemented the static tests mentioned above
with the use of the SNM. An example of the outcome of this test is reported
focusing only in the NCOR system.

The general conclusion that stems from the dynamic tests is that an
interaction between species that exclusively prefers phylogenetic proximity\
can never give rise to a nested contact pattern. A nested contact pattern
turns out to be unstable in the presence of such interaction rule. A way to
see this is by realizing that such interaction mechanism relies in the
generalized occurrence of species that are specialists therefore ruling out
generalists. However, these are an indispensable ingredient of a nested
organization. A contact rule governed by phylogenetic proximity for \textit{%
both mutualists guilds} therefore tends to destroy a nested pattern of
contacts. These tests indicate in addition that the adjacency matrices would
tend to break down into separate, nearly independent components in which
groups of phylogenetically close neighbors of both guilds hold contacts
among each other but not with the rest of the species of the ecological
system. The stability analysis performed using the SNM proves that a contact
pattern such as the ideal PERF system is unstable under the perturbations
produced by a contact rule favoring phylogenetic proximity.

We have also proved that an alternative interaction pattern dominated by 
\textit{phylogenetic diversity} is instead a much better approach to
describe real situations. This interaction mechanism is one in which species
are assumed to hold contacts with counterparts that are already visited by a
greater \textit{diversity} of species. This could be interpreted as a
stylized version of a rule by which all species tend to put the least
possible requirements in their feeding or pollinizing counterparts. It is
worth to stress that this rule is fully consistent with the ones tested in
Refs.\cite{nosotros1} and \cite{nosotros2} and provides an additional proof
of the validity of the possible sources of nestedness mentioned there. In
those situations highly realistic degree distribution functions and contact
patterns are produced by the SNM by only assuming that species tend to hold
contacts with species that already hold a greater number of contacts.

The second contact rule by which one species tends to visit a set of
counterparts having a maximal diversity agrees in fact with one of the two
schemes suggested in the opening figure of Ref.\cite{Rezende}. In addition
if a group of phylogenetically close species happen to have similar contact
patterns, the corresponding contact pattern turns out to be stable under
such maximal diversity interaction rule. A set of phylogenetically close
species of the NCOR system that also are good generalists is stable under
the perturbations of the SNM. This points in the direction that, although
phylogenetic affinity can not determine the existence of nestedness it is
fully \textit{compatible} with it.

The results obtained with the dynamic tests are consistent with those
obtained with the static ones. Both point into the same direction of
considering the correlation of phylogenetic proximity and degree, if any, is
largely accidental and that it can never be considered as a general cause of
nestedness. A dominant cause of the generalized nestedness found in
mutualistic ecosystems perhaps lies on the simple fact that species that we
observe in real systems today are those that tend to put the least possible
restrictions on their mutualist counterparts.

\section*{Acknowledgments}

The authors wish to acknowledge helpful discussions and criticism from D.
Medan and M. Devoto.


\begin{thebibliography}{99}
\bibitem{nota} This is the only order that allows to see some structure in
the ATC of nested systems. In the limit of perfect nestedness, an attack in
increasing order only shows its effect when the last species is eliminated.
\end{thebibliography}
\end{document}